\documentstyle[epsfig]{aipproc}

\newcommand{\he}{{\mathrm H}{\mathrm e}}
\newcommand{\neu}{{\mathrm n}}
\newcommand{\pro}{{\mathrm p}}
\newcommand{\deu}{{\mathrm d}}
\newcommand{\ele}{{\mathrm e}}

\begin{document}

\title{Neutrino Heating in an Inhomogeneous Big Bang Nucleosynthesis Model}

\author{Juan F. Lara$^*$}
\address{$^*$Center for Relativity, University of Texas at Austin\\
         Austin, Texas 78712}

\maketitle

\begin{abstract}

The effect of the heating of neutrinos by scattering with electrons and 
positrons and by $e^{-}e^{+}$ annihilation on nucleosynthesis is calculated for
a spherically symmetric baryon inhomogeneous model of the universe.  The model
has a high baryon density core and a low density outer region.  The heating 
effect is calculated by solving the Boltzmann Transport Equation for the 
distribution functions of electron and muon/tau neutrinos.  For a range of 
baryon-to-photon ratio $\ln ( \eta_{10} ) = [ 0, 1.5 ]$ and 
$ r_{i} = [ 10^{2}, 10^{8} ]$ cm the heating effect increases the mass fraction
$X_{ ^{4}\he}$ by a range of $\Delta X_{ ^{4}\he} = [1, 2] \times 10^{-4}$. 
The change of the value of $X_{ ^{4}\he}$ appears similiar to the change
caused by an upward shift in the value of $\eta_{10}$.  But the change to 
deuterium is a decrease in abundance ratio $Y(\deu)/Y(\pro)$ on the order of 
$10^{-3}$, one order less than the decrease due to a shift in $\eta_{10}$.  

\end{abstract}

\section{Introduction}

When discrepencies arise between observations of light isotope abundances and
the predictions of standard ( homogeneous and isotropic ) Big Bang 
Nucleosynthesis (BBN ) models research has turned to BBN models with 
inhomogeneous baryon distributions ( IBBN models ) to attempt to resolve the 
discrepencies.  Articles in the 1970's and 1980's looked for IBBN models with 
an overall baryon density that both equalled the critical density and satifised
observational constraints on the light elements. \cite{MM:1993} These IBBN 
models could not satisfy observational constraints to isotope abundances for 
baryon densities other than densities demanded by standard models.  In the late
1990's observations of $ ^{4}$He and deuterium placed conflicting constraints
on the baryon density in standard models.  ( KKS, 1999 ) \cite{KKS:1999} found
a range of agreement for the outer fringes of the $2 \sigma$ ranges of the
observations using an IBBN model code.  Most recently ( KS, 2000 )
\cite{KS:2000} used an IBBN model to bring baryon density constraints from 
cosmic microwave background observations in agreement with constraints from
$ ^{4}$He and deuterium observations, though not with $ ^{7}$Li observations.

Nucleosynthesis models have become more descriptive as they include smaller 
effects that faster computers can calculate accurately.  One effect is the 
neutrino heating effect.  Electrons and positrons pass a fraction of their 
energy to neutrinos through annihilation and scattering.  That energy transfer
can change the rates of neutron-proton interconversion, and then the results of
nucleosynthesis.  Hannestad and Madsen \cite{HM:1995} derive a means of solving
the Boltzmann Transport Equation for the neutrino distribution functions, 
presenting results for a standard BBN model.  This author will discuss the 
neutrino heating effect in an IBBN model.  This article will show how the 
heating effect slightly alters neutron distribution and nucleosynthesis, and 
show the effect's dependence on IBBN parameters.  

\section{The Model}

The IBBN code for this article corresponds to a model with a spherically 
symmetric baryon distribution.  The model is divided into a core and 31 inner
shells with high baryon density, and 32 outer shells with low baryon density.
A run starts at electromagnetic plasma temperature $T = $ 100 GK.  The distance
scale $r_{i}$, the radius of the model when $T = $ 100 GK, can be varied.
During a run the number density $n(i,s)$ of isotope species $i$ in shell $s$ is
determined by the equation \cite{MMAF:1990}

\begin{eqnarray}
   \frac{dn(i,s)}{dt} & = & \frac{1}{n_{b}(s)} \sum_{j, k, l} N_{i} \left
     ( - \frac{n^{N_{i}}(i,s) n^{N_{j}}(j,s)}{N_{i}! N_{j}!} [ij] + 
              \frac{n^{N_{k}}(k,s) n^{N_{l}}(l,s)}{N_{k}! N_{l}!} [kl] \right )
                            \nonumber \\
                      &   & - 3 \dot{\alpha} n(i,s) + \frac{1}{r^{2}}
    \frac{\partial}{\partial r} \left ( r^{2} D_{n} 
                        \frac{\partial n(i,s)}{\partial r} \right ) 
\end{eqnarray}

\noindent The first term corresponds to nuclear reactions and beta decays 
within shell $s$, the second term to the expansion of the universe, and the 
third term to diffusion of isotope $i$ between shells.  In this model only 
neutrons diffuse.  As $T$ falls neutrons diffuse from the high density shells 
to the low density shells, until the neutrons are homogeneously distributed.
The weak reactions n + $\nu_{\ele} \leftrightarrow$ p + e$^{-}$, n + 
e$^{+} \leftrightarrow$ p + $\bar{\nu_{\ele}}$, and n $\leftrightarrow$ p + 
e$^{-} + \bar{\nu}_{\ele}$ convert protons in the high density shells to 
neutrons, which then diffuse into the low density shells where the weak 
reactions convert them back to protons.  Nucleosynthesis occurs earlier in the
high density shells, depleting neutrons.  Neutrons from the low density shells 
then back diffuse into the high density shells until nucleosynthesis 
incorporates all neutrons into nuclei, mostly $ ^{4}$He nuclei.

\begin{figure}
   \leftline{\rotatebox{270}{\resizebox{3in}{3in}
                             {\includegraphics{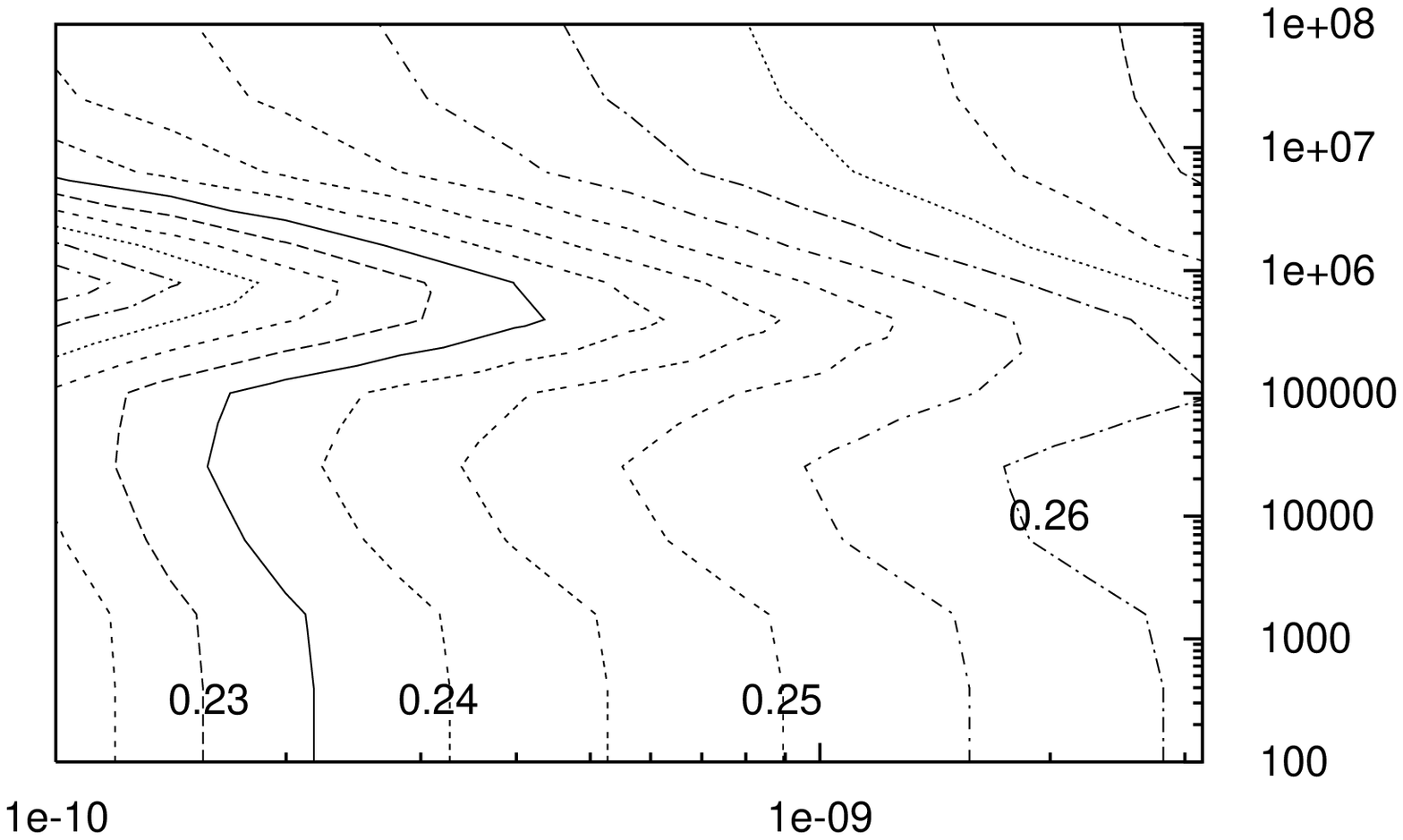}}}
             \rotatebox{270}{\resizebox{3in}{3in}
                             {\includegraphics{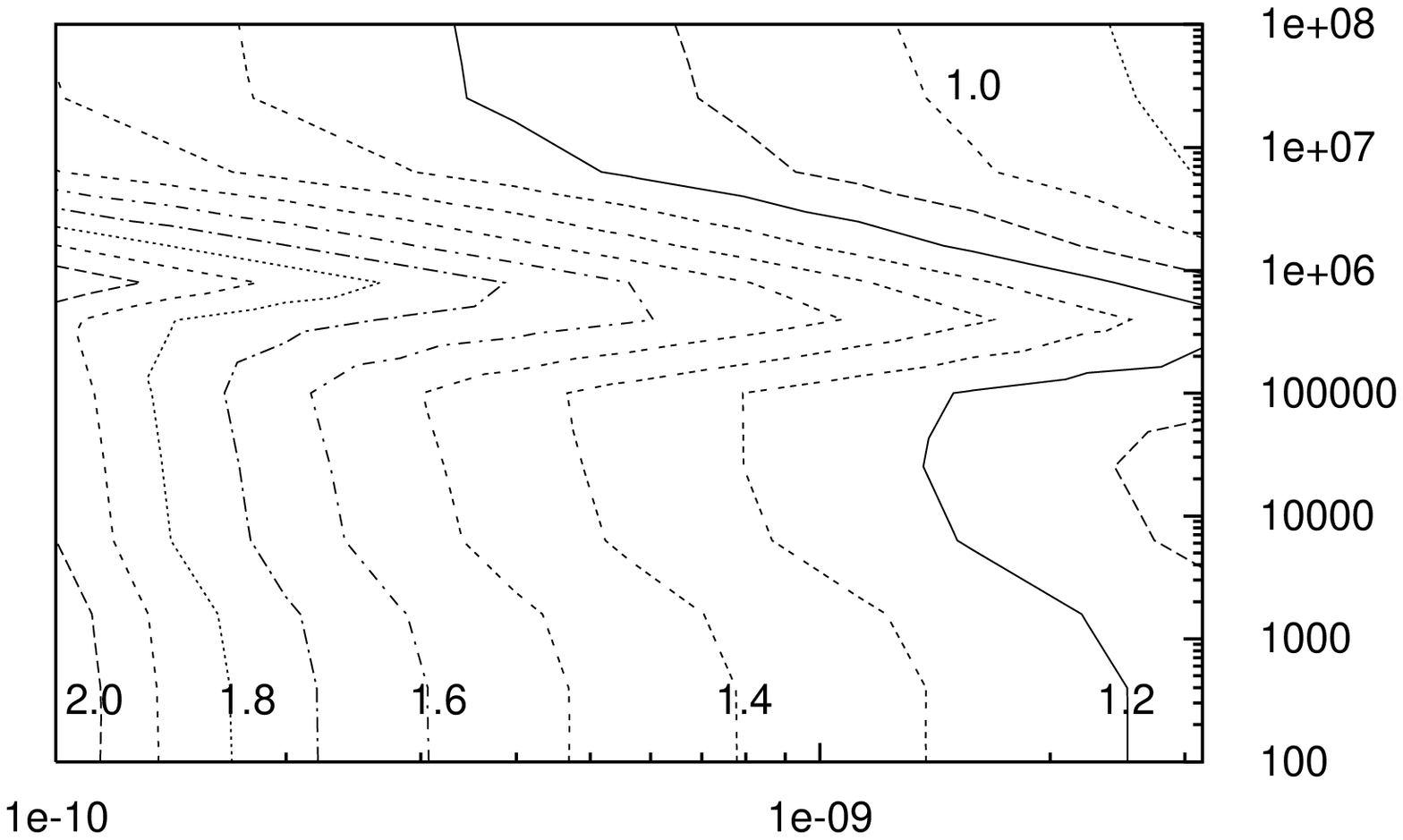}}}}
   \caption{$X_{ ^{4}\he}$ (1a) and $10^{4} \Delta X_{ ^{4}\he}$ (1b) due to 
            neutrino heating.}
\end{figure}

Figure (1a) shows the mass fraction $X_{ ^{4}\he}$ of $ ^{4}$He nuclei produced
by the model, as a function of $\eta$ and $r_{i}$ ( in centimeters ).  Contour
lines correspond to specific values of $X_{ ^{4}\he}$.  Neutron diffusion 
occurs later as $r_{i}$ increases.  The shape of the contour lines is 
determined by when neutron diffusion occurs compared with the weak reaction
rates mentioned above and with nucleosynthesis.  As $r_{i}$ increases the time
of neutron diffusion coincides less and less with the time before the weak 
reactions fall out of equilibrium.  More protons then remain in the high 
density shells.  When $r_{i} = $ 25000 cm the innermost shells retain all their
protons, and nucleosynthesis occurs earlier enough that nearly all neutrons 
back diffuse into the high density shells to undergo nucleosynthesis.  
$X_{ ^{4}\he}$ increases as $r_{i}$ increases to 25000 cm.  But for larger 
$r_{i}$ neutrons cannot reach the innermost shells before nucleosynthesis, 
leading to decreasing production of $ ^{4}$He.  $ ^{4}$He production bottoms 
out at $r_{i} = $ 790000 cm when diffusion occurs at the same time as 
nucleosynthesis.  For higher $r_{i} = $ the high density shells and the low 
density shells act separately from each other, with the high density shells 
producing large amounts of $ ^{4}$He.

To calculate the neutrino heating effect, the model solves the Boltzmann 
Transport Equation \cite{HM:1995}

\begin{eqnarray}
   \frac{d f_{1}(p_{1})}{dt} 
               & = & \frac{1}{2 E_{1}} \left ( \frac{kT_{N}}{R} \right )^{5}
                     \int \frac{d^{3} p_{2}}{( 2\pi )^{3} 2E_{2}}
                     \int \frac{d^{3} p_{3}}{( 2\pi )^{3} 2E_{3}} \nonumber \\
               &   & \int \frac{d^{3} p_{4}}{( 2\pi )^{3} 2E_{4}} S|M|^{2} 
         ( 2\pi )^{4} \delta^{4} ( p_{1} + p_{2} - p_{3} - p_{4} ) \nonumber \\
               &   & \{ [ 1 - f_{1}(p_{1}) ] [ 1 - f_{2}(p_{2}) ] 
                  f_{3}(p_{3}) f_{4}(p_{4}) \nonumber \\
               &   & - f_{1}(p_{1}) f_{2}(p_{2}) [ 1 - f_{3}(p_{3}) ]
                     [ 1 - f_{4}(p_{4}) ] \} 
\end{eqnarray}

\noindent for the neutrino distribution functions $f_{i}(p_{i})$.  The 
distribution functions are used to calculate the neutrino energy densities, and
$f_{\nu_{\ele}}(p_{\ele})$ for electron neutrinos is used to calculate the 
neutron-proton conversion reaction rates.  Increased $f_{\nu_{\ele}}(p_{\ele})$
makes the conversion reactions produce fewer neutrons for nucleosynthesis, but
decreased electron energy density $\rho_{\ele}$ makes those reactions produce
more neutrons at the same time.  Decreased $\rho_{\ele}$ decreases temperature
$T$, making nucleosynthesis occur earlier, when more neutrons are present.
This last effect, the clock effect, ultimately determines the increase in
$X_{ ^{4}\he}$ due to neutrino heating.

\section{Results}

The diffusion coefficient $D_{\neu}$ is calculated from collisions between 
neutrons and electrons/positrons ( $D_{\neu\ele}$ ) and neutrons and protons 
( $D_{\neu\pro}$ ). \cite{KAGMBCS:1992}

\begin{eqnarray}
   \frac{1}{D_{\neu}} & = & \frac{1}{D_{\neu\ele}} + \frac{1}{D_{\neu\pro}} \\
   D_{\neu\ele} & = & \frac{3}{8} \sqrt{\frac{\pi}{2}} \frac{c}{n_{\ele}
                      \sigma_{\neu\ele}} \frac{K_{2}(z)}{\sqrt{z} K_{5/2}(z)}
                      \left ( 1 - \frac{n_{\neu}}{n_{t}} \right ) \\
              z & = & \frac{m_{\ele}}{kT} \nonumber
\end{eqnarray}

\begin{figure}
   \leftline{\resizebox{3in}{3in}{\includegraphics{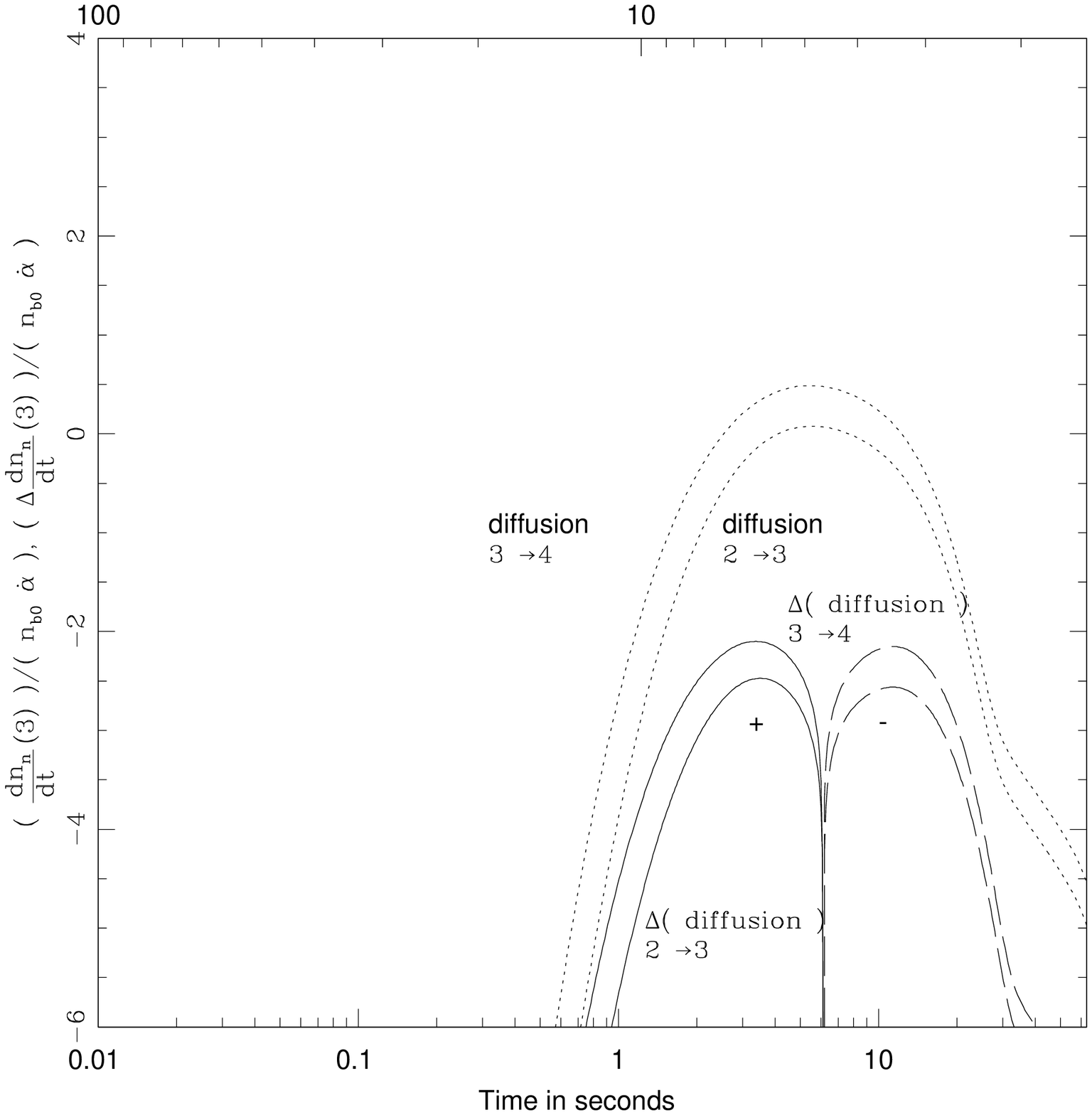}}
             \resizebox{3in}{3in}{\includegraphics{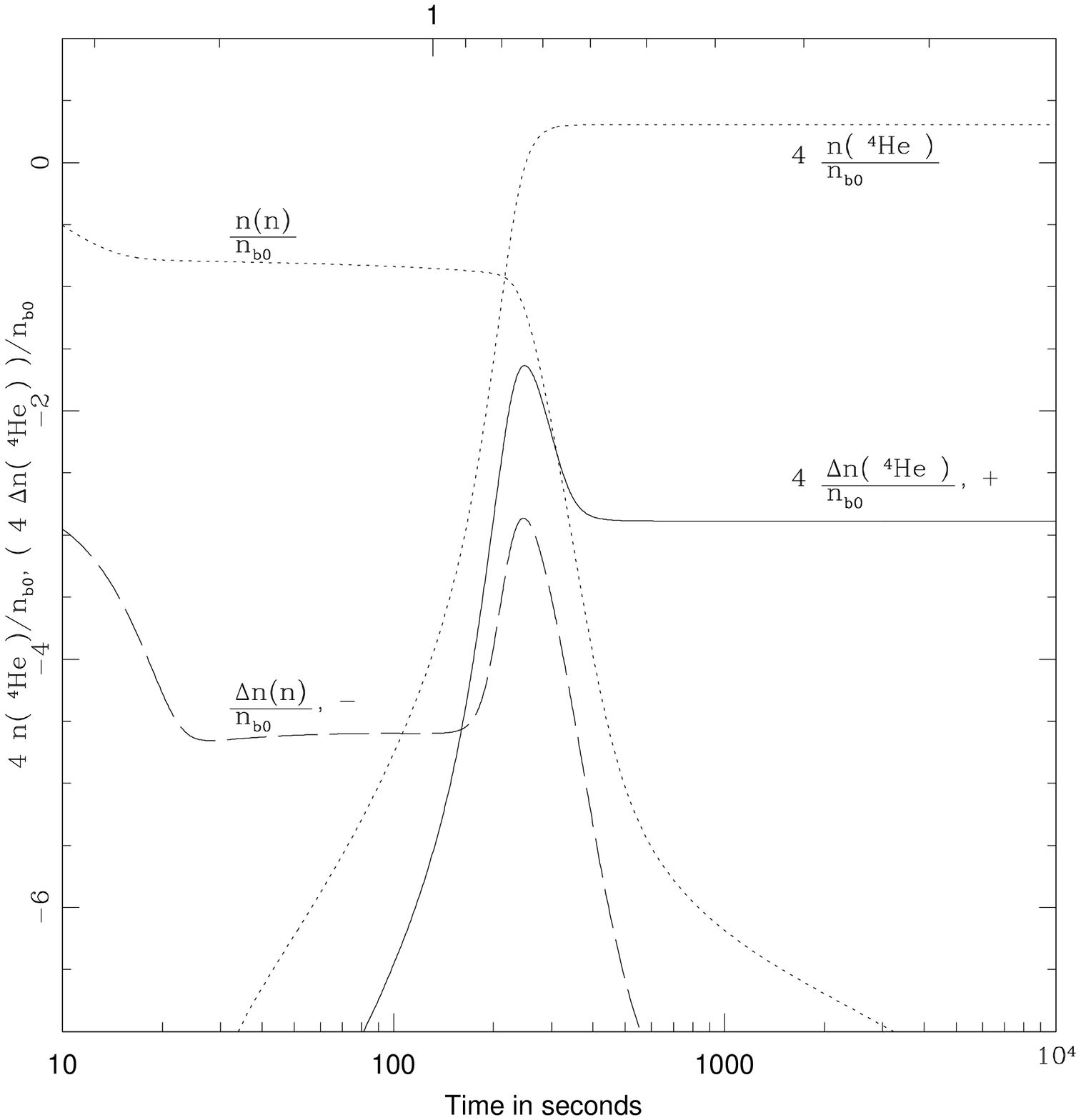}}}
   \caption{In Figure (2a) the dotted lines are the diffusion rates to and from
            shell 3, and the solid ( increase ) and dashed ( decrease ) lines 
            are the changes of the rates due to neutrino heating.  Figure (2b)
            shows $n(\neu)$ and $n( ^{4}\he)$ ( dotted lines ) and the changes
            in these number densities due to heating.  $n_{bO}$ is a 
            normalizing factor and $\dot{\alpha}$ is the expansion rate of the
            universe.}
\end{figure}

Neutrino heating affects $D_{\neu\ele}$ by lowering the electron number density
$n_{\ele}$ and lowering the electromagnetic plasma temperature $T$ at a given 
time.  The neutrinos and the electrons/positrons are homogeneously distributed
throughout this model. So $D_{\neu\ele}$ is the same for all the shells. 

Figure (2a) shows the neutron diffusion rates into and out of Shell 3, a high 
density inner shell, for the case of $\eta_{10} = $ 3.0 and $r_{i} = $ 25000 
cm.  Figure (2a) also shows the change in the rates due to heating.  This 
change is at first positive, and then becomes negative.  This change 
corresponds to a shift in the diffusion rate to a time earlier by 0.1 \%.  All
other shells also have this earlier time shift.  Figure (2b) shows the number 
densities of neutrons ( $n_{\neu}$ ) and $ ^{4}$He nuclei ( $n_{ ^{4}\he}$ ) at
around $T = $ 0.8 GK, when nucleosynthesis occurs, for shell 3 and the same 
values of $\eta_{10}$ and $r_{i}$.  Figure (2b) also shows the changes in 
$n_{\neu}$ and $n_{ ^{4}\he}$ due to neutrino heating.  The clock effect
mentioned above is seen in the humps in the graphs for the changes 
$\Delta n_{\neu}$ and $\Delta n_{ ^{4}\he}$.  
$\Delta n_{\neu}$ and $\Delta n_{ ^{4}\he}$ are larger between the time when 
neutrons are converted to $ ^{4}$He nuclei with the heating effect on and when
neutrons are converted to $ ^{4}$He nuclei without heating.

Figure (1b) shows $10^{4} \Delta X_{ ^{4}\he}$, the change due to neutrino
heating.  For the ranges of $\eta$ and $r_{i}$ observed $\Delta X_{ ^{4}\he}$
remains within a range of $[ 1.1, 2.0 ] \times 10^{-4}$.  $X_{ ^{4}\he}$ is
determined by the baryon number densities in the shells, and those densities 
also determine the magnitude of the clock effect.  That seems to account for
the contour lines of $\Delta X_{ ^{4}\he}$ tracking the contour lines of 
$X_{ ^{4}\he}$ itself.  The heating effect on $X_{ ^{4}\he}$ seems to be 
similar to the effect one would get by shifting the value of $\eta$ upwards by
about 1 to 2 \%.  The earlier shift in neutron diffusion due to heating would
have the effect of the lines in Figure (2a) being stretched to higher values of
$r_{i}$.  So for a given value of $r_{i}$ and $\eta$ the change 
$\Delta X_{ ^{4}\he}$ will have a value corresponding to $X_{ ^{4}\he}$ at that
point in the graph plus a shift corresponding to a lower value of $r_{i}$.  
The change is bigger for $r_{i} \le $ 25000 cm, when $X_{ ^{4}\he}$ decreases
with decreasing $r_{i}$, and is smaller for 25000 cm $ < r_{i} \le $ 790000
cm when $X_{ ^{4}\he}$ increases with decreasing $r_{i}$.

Figure (3a) shows the overall abundance ratio $Y(\deu)/Y(\pro)$ between 
deuterium and protons.  Figure (3b) shows 
$10^{3} \Delta \log [ Y(\deu)/Y(\pro) ]$, the change due to neutrino heating.
The nuclear reaction rates that create and destroy deuterium depend on the 
electromagnetic plasma temperature $T$ and on shell $s$'s baryon energy density
$\rho_{b}(s)$.  Through $T$ neutrino heating shifts the nuclear reaction rates
to a slightly earlier time.  So for $r_{i} < 10^{5}$ cm less deuterium remains
at the end of a run.  This decrease in $Y(\deu)/Y(\pro)$ due to heating turns
out to be one order of magnitude less than the decrease that would come from 
an increase in $\eta_{10}$ mentioned above.  For $\eta_{10} = $ 3.0 and 
$r_{i} = $ 25000 cm one has to increase $\eta$ by 1.53 \% to increase 
$X_{ ^{4}\he}$ by as much as neutrino heating does.  But that upward shift 
would decrease $\log [ Y(\deu)/Y(\pro) ]$ by around 9.19 $\times 10^{-3}$,
instead of the decrease of 1.12 $\times 10^{-3}$ done by heating.  For
$r_{i} > 10^{5}$ cm the low density outer shells produce considerable amounts
of deuterium.  In that range the heating effect increases deuterium production.
The neutrino heating effect on $ ^{4}$He can then be distinguished from a shift
in $\eta_{10}$ by looking at how $Y(\deu)/Y(\pro)$ differs from the results 
without heating.

\begin{figure}
   \leftline{\rotatebox{270}{\resizebox{3in}{3in}
                             {\includegraphics{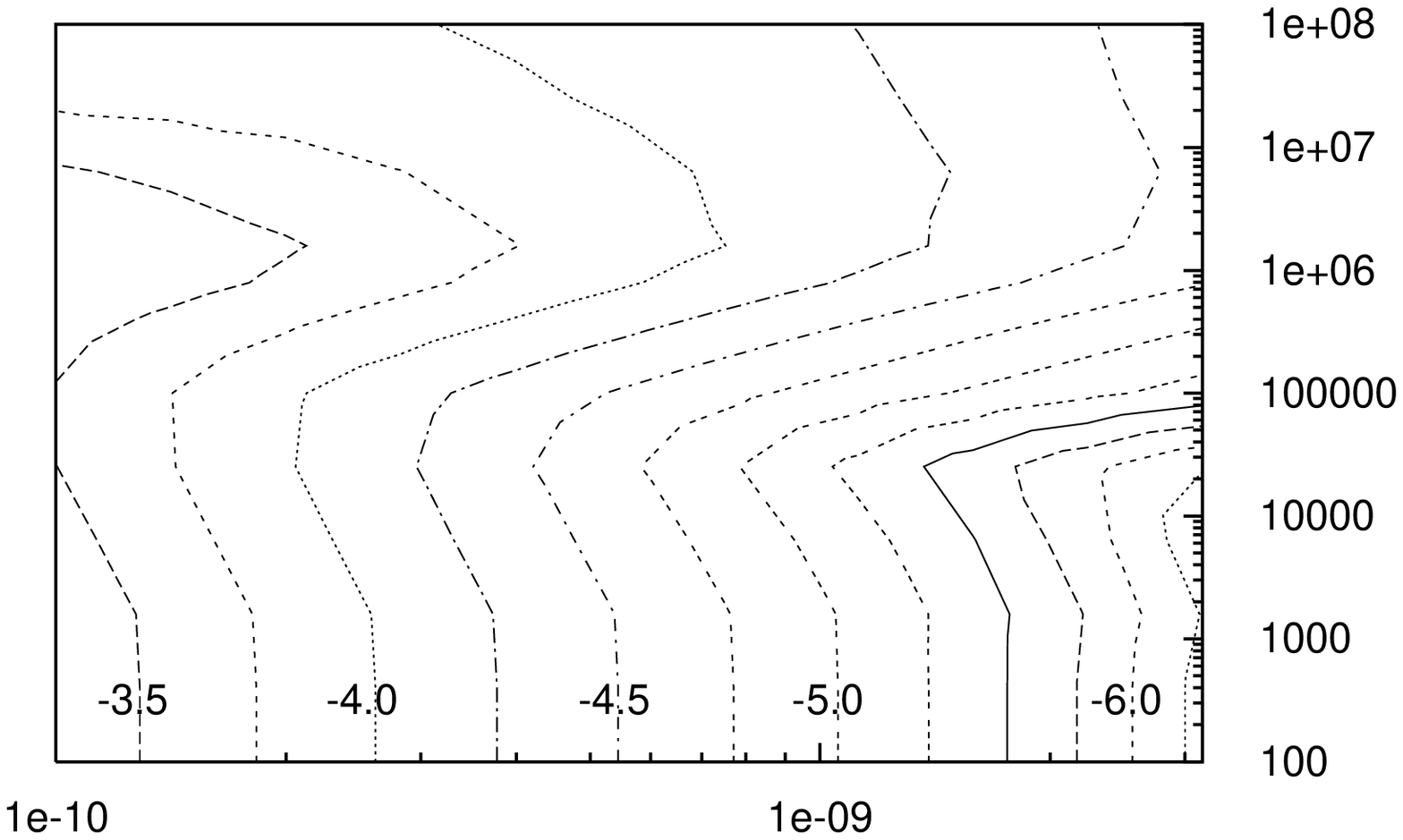}}}
             \rotatebox{270}{\resizebox{3in}{3in}
                             {\includegraphics{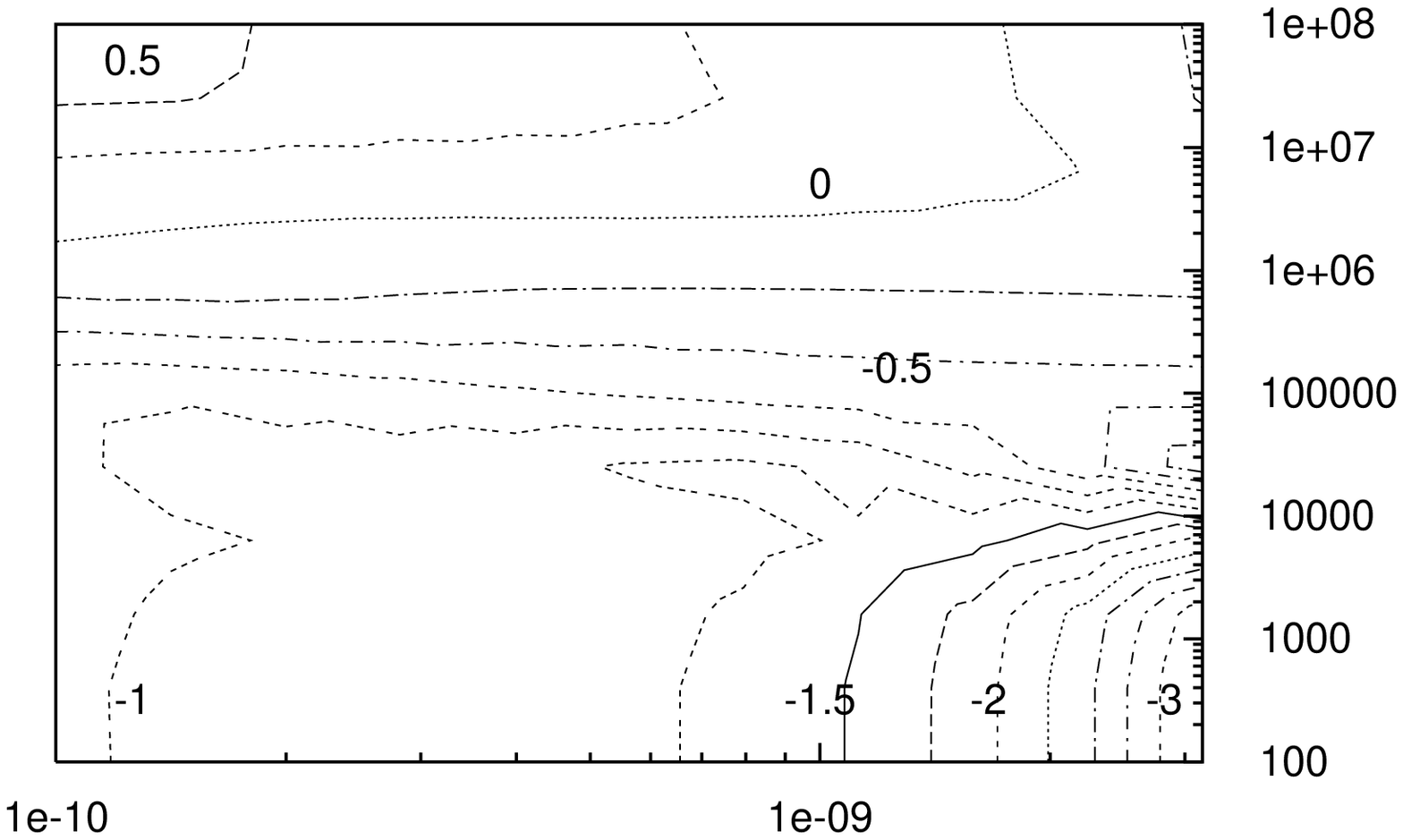}}}}
   \caption{$\log [ Y(\deu)/Y(\pro) ]$ (3a) and 
            $10^{3} \Delta \log [ Y(\deu)/Y(\pro) ]$ (3b) due to neutrino 
            heating.}
\end{figure}

\section{Conclusions and Future Research}

The neutrino heating effect increases the mass fraction of $ ^{4}$He by 
$[ 1.1, 2.0 ] \times 10^{-4}$ for the observed range of baryon-to-photon 
ratio $\eta_{10} = [ 1, 32 ]$ and distance scale $r_{i} = [ 10^{2}, 10^{8} ]$
cm.  That's the same order of magnitude as in the standard case.  The increase
in the mass fraction resembles the increase that would come from increasing 
$\eta_{10}$ by about 1 \% to 2 \%.  The heating effect also decreases 
deuterium production.  But the decrease in deuterium is an order of magnitude 
less than the decrease that would come from the increase of $\eta_{10}$ 
mentioned above.  Further, the effect increases deuterium production for 
$r_{i} \ge 10^{5}$ cm.

A larger future article can show the earlier time shift of neutron diffusion 
for values of $\eta_{10}$ and $r_{i}$ other than 3.0 and 25000 cm, and for
shells other than shell 3 as well.  These graphs would show how this time 
shift is characteristic of the whole observed range of $\eta_{10}$ and $r_{i}$.
The article can also look at deuterium production for $r_{i}$ lower than and
higher than $10^{5}$ cm, to explain in more detail how the change due to 
neutrino heating changes signs.  The article can also comment on production of
other isotopes, such as $ ^{7}$Li.  One can determine if the change due to 
heating has a distance scale dependence similar to the dependence for 
deuterium.

One should note that CMB observations have placed parameters on the 
baryon-to-photon ratio ( $\Omega_{B} h^{2} = 0.032_{-0.008}^{+0.009}$ where
$\Omega_{B} h^{2} = ( 3.650 \pm 0.008 ) \times 10^{-3} \eta_{10}$ ) 
\cite{BNT:2000}.  A future article should particularly look at the IBBN code's
behavior in regions corresponding to these parameters.

\end{document}